\documentclass[Transaction, 9pt]{IEEEtran}
\usepackage{times}
\usepackage{graphicx}
\usepackage{eqnarray}
\usepackage{amsmath}
\usepackage{setspace}
\usepackage{gfcoeff}
\usepackage{cite}
\begin{document}
\title{ Electrostatically Doped Heterojunction TFET with Enhanced Driving Capabilities for Low Power Applications}
\author{Kanchan~Cecil and Jawar~Singh* 
\thanks{*The authors are with the Department of Electronics and Communication Engineering, PDPM Indian Institute of Information Technology Design and Manufacturing, Jabalpur,
MP, 482005 India. E-mail: (c.kanchan@iiitdmj.ac.in; jawar@iiitdmj.ac.in).}}
\maketitle
\begin{abstract}
This paper projects the enhanced drive current of a n-type electrostatically doped (ED) tunnel field-effect transistor (ED-TFET) based on heterojunction and band-gap engineering via TCAD 2-D device simulations. The homojunction ED-TFET device utilizes the electrostatic doping in order to create the source/drain region on an intrinsic silicon nanowire that also felicitates dynamic re-configurability. The ED-TFET offers good electrostatic control over the channel with reduced thermal budget and process complexity. However, device exhibits low ON current, therefore, in this work, we elaborate on interfacing of group III-V with group IV semiconductors for heterojunction. Incorporation of heterojunction and band gap engineering in the ED-TFET has improved  drive current even at very low operating voltage. The comparison of various low band gap source region materials shows that germanium (Ge) source (Si-Si-Ge) ED-TFET provides steepest subthreshold swing (SS) of about 9.5 mV/dec, and higher ON-state drive current of 1.58 mA at $V_{DS}$ = 1 V and 0.093 mA at $V_{DS}$ = 0.5 V with same SS.
\end{abstract}


\section{Introduction}
Continuous downscaling of CMOS technology has significantly improved the performance, functionality, and packaging density, at the cost of rise in power dissipation and complex fabrication process of nanoscale CMOS devices. The rise in power dissipation is due to faster switching activities and subthreshold leakage. In this pursuit, small subthreshold swing SS (sub-KT/q) devices have acquired wider attention due to their efficient switching and very low subthreshold leakage. Among the emerging sub-kT/q devices, tunnel field-effect transistors (TFETs) have considered as the potential devices on account of their low SS, low leakage current and integration compatibility with CMOS process~\cite{Riel, Woo, Boucart}. However, TFETs have comparatively low ON-state current as compared to their counterpart MOSFET devices due to poor band-to-band tunneling phenomenon (BTBT). The efficient tunneling in TFETs requires abruptly doped junctions which enforce complex fabrication process with high thermal budget requirements due to ion-implantation and thermal annealing techniques. Apart from steep doping requirements, these junctions restrain aggressive scaling and enhanced random dopant fluctuations (RDFs), as a result, these devices become susceptible to process variations related issues~\cite{Changhwan}.

Therefore, before considering the potential merits of TFETs, issues related to steep doping requirements at drain/source junctions, process variability and complex fabrication process need to be addressed. In this line, an electrostatically doped (ED) TFET was proposed that attempted to address the aforementioned issues~\cite{Sahu}. The ED-TFET employs an intrinsic silicon nanowire for the formation of a drain (D) -- channel -- source (S) structure without externally doped S/D regions. For making the $p^{+}$ and $n^{+}$ S/D regions, the polarity gate (PG) concept was employed in which appropriate bias at polarity gates were applied, as a result, desired level and type of carriers can be induced. This process of inducing carriers in a intrinsic silicon nanowire is referred as electrostatic doping~\cite{Marchi}. The ED eliminates requirements of external doping, thereby yields, low thermal budget and simplified fabrication process. Further, intrinsic nature of silicon nanowire provides less susceptibility towards process parameter and temperature variations~\cite{Sahu}. However, ED-TFET faces an inherent problem of poor ON current which makes it less suitable for main stream VLSI circuit applications.

\begin{figure} 
\center
\includegraphics[width=80mm, keepaspectratio]{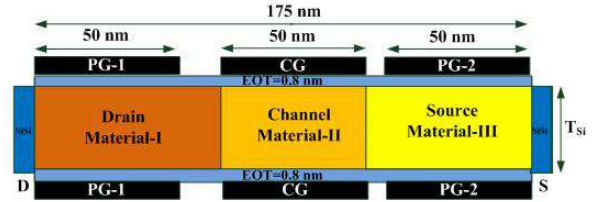}
\caption{Cross-sectional view of an electrostatically doped (ED) TFET~\cite{Sahu}.}
\label{strun}
\end{figure}

The poor ON current issue of TFETs has been addressed through many ways in literature, such as, employing germanium (Ge) channel~\cite{Royer}, strained silicon~\cite{Toh}, and group III-V channel and hetero-junction structure~\cite{Yoon, Ghosh}. Apart from that band gap engineering has also been explored for the selection of materials at drain and source sides for optimal drive current as well as $I_{ON}$/$I_{OFF}$ ratio. Recently, many successful attempts have been made in order to fabricate such devices with hetero-junctions through epitaxial growth with vertical interface (or vertical hetero-structures)~\cite{Chao,Zhou}. Although all these approaches have yielded enhanced ON current in TFETs, yet the cost, fabrication complexity, and process variations issues still need to be addressed.

In this paper, poor ON current issue of the ED-TFET  is explored through hetero-junction and band-gap engineering, while retaining it`s inherent merits of reduced thermal budget requirements, fabrication complexity, and less susceptibility to process variation. For ON current enhancement of this ED-TFET, we have explored interfacing of semiconductor materials (group III-V and group IV)  with different combinations employing band gap engineering  at source-channel as well as at drain-channel junctions. From the simulation results, it is confirmed that the combination of group III-V and group IV and also group IV  (Si-Ge) materials for hetero source-channel interface may exhibit higher ON current as well as low OFF current.

\section{Device Structure and Simulation Parameters}

Fig.\ref{strun} show the cross-sectional view of a silicon nanowire based ED-TFET structure~\cite{Sahu}. In this device structure, drain (D) -- channel -- source (S) regions have been marked with three different materials, later combination of these materials for different regions have been optimized for enhanced ON current, low OFF current, and other performance parameters. The simulation parameters considered are as follows~\cite{Sahu}: ultrathin silicon film thickness ($T_{si}$) = 10 nm, equivalent  oxide thickness (EOT) = 0.8 nm, channel length ($L_{g}$) = 50 nm, spacing between gate and source electrode $L_{Gap,S}$ = 3 nm, and spacing between gate and drain electrode $L_{Gap,D}$ = 15 nm. The metal control gate work function, $WF_{CG}$ of 4.5 eV of ED-TFET has been optimized to 4.3 eV. $WF_{CG}$=4.3 eV  causes a depletion region in the channel in the OFF state. The simulated device is a dopingless architecture with intrinsic body having uniform doping concentration of $N_{D}$ = $10^{15}$. The formation of drain(D)-channel-source(S) is achieved with the appropriate bias at polarity gates (PGs) (i.e electrically doped) yield  $p^{+}$ and $n^{+}$ S/D regions. Further, this electrostatic doping can also be exploited for dynamic configurability to ED-TFET to switch between n- and p-type TFETs.

The interfacing of low and high band gap materials (group III-V and group IV semiconductor materials) results in heterojunctions in the ED-TFET, therefore, the ED-TFET has junction between two semiconductors instead of metallurgical junctions in conventional TFETs. The materials for drain--channel and source -- channel heterojunctions were chosen according to their band gaps for higher ON current and low OFF current. The low band gap materials, such as Germanium (Ge), Silicon-Germanium ($Si_{0.3}$$Ge_{0.7}$), and Gallium Antimonide (GaSb) can be employed for source region, whereas, Indium Phosphide (InP) can be used as a high band gap material for drain and channel region.

The ED-TFET device is simulated using a 2-D device simulator, Atlas Silvaco V5.19.20 \cite{Atlas}. A nonlocal band-to-band tunneling model is used for the simulation of tunnel current. This inter band tunneling current in the TFET depends on the potential profile along the entire path connected by tunneling. Therefore, nonlocal models uses Wentzel-Kramer-Brillouin approximation method for estimating the tunneling probability along the tunneling path. The results are obtained through  drift-diffusion current transport model where the poisson and carrier continuity equations are solved self consistently. Concentration dependent Shockley-Read-Hall (SRH) generation and recombination model is incorporated for transitions that occur in the presence of traps or defects. The tunneling model was first calibrated by reproducing the results reported in~\cite{Hwan} in order to avoid the use of default parameters of the model for Ge. The fitted coefficients A = 1.53x$10^{22}$ $cm^{-3}$.$s^{-1}$ and B = 1.38x$10^{8}$ V.$cm^{-1}$ were used throughout in our simulation work.


\section{Band diagrams and Device Characteristics}
\begin{figure} 
\center
\includegraphics[width=85mm,keepaspectratio]{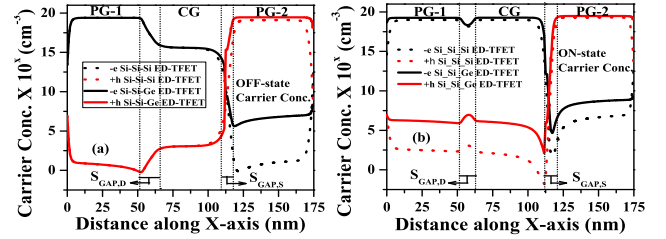}
\caption{Carrier concentration (a) OFF state ($V_{CG}$=0V and $V_{DS}$=1V), and (b) ON state ($V_{CG}$=1.2V and $V_{DS}$=1V) for both Si and Ge Source ED-TFETs.}
\label{CC_Comp}
\end{figure}

Fig.\ref{CC_Comp} (a) and (b) show the simulated carrier concentration of induced electrons and holes in the ED-TFETs with for different source materials (silicon and germanium) under OFF-state and ON-state along a horizontal cut-line at 1 nm below the Si/$SiO_{2}$ interface. The profile of both holes and electrons induced through electrostatic doping (i.e. appropriate PGs biases $V_{PG-1}$=1.2 V and $V_{PG-2}$=-1.2 V), is similar with respect to conventionally doped TFETs except a sudden drop at nickel-silicide S/D contacts. It can be inferred that the induced carrier concentration is higher in case of germanium source ED-TFET due to low work function of Ge. In OFF state ($V_{CG}$=0V), the intrinsic channel under the control gate is depleted  due to sufficient work function differences between metal (4.3 eV) and semiconductor, and therefore, causes almost no current. While under ON-state ($V_{CG}$=1.2V), intrinsic region under the control gate becomes $n^{+}$ having electron concentration of the order of $10^{19}$ $cm^{-3}$ and the device works on flat-band condition with almost zero resistance. This ensures formation of an abrupt p-n junction at source-channel interface necessary for tunneling of carriers from valence band (VB) of source to conduction band (CB) of channel.

\begin{figure} 
\center
\includegraphics[width=85mm,keepaspectratio]{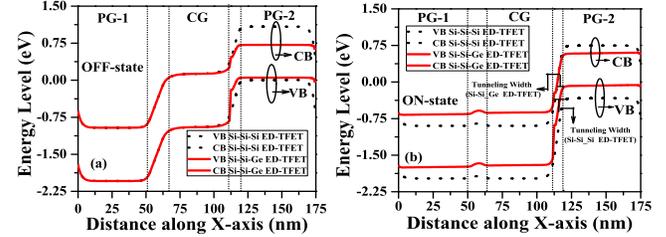}
\caption{Energy band diagrams (a) OFF state ($V_{CG}$=0V and $V_{DS}$=1V), and (b) ON state ($V_{CG}$=1.2V and $V_{DS}$=1V) for Si and Ge Source ED-TFETs.}
\label{EB_Comp}
\end{figure}

Fig.\ref{EB_Comp} (a) and (b) show the valance band (VB) and conduction band (CB) energies of Si as well as Ge source ED-TFETs in OFF state as well as in the ON state. In OFF state, it can be seen that tunneling probability of electron from VB of source to CB of channel is low due to higher energy barrier width at source-channel junction of Si source ED-TFET in comparison with Ge source ED-TFET. Comparatively less tunneling width of Ge source ED-TFET is due to the low band gap of germanium with respect to Si which results in significant BTBT tunneling in OFF state also and hence higher OFF-state current. On application of positive gate bias (ON state), the CB and VB in the channel region gets aligned with CB and VB of drain region, causes the further reduction in tunneling width in case of Ge source ED-TFET, and thereby, increases the probability of electrons tunneling from VB of source to CB of channel. Therefore, Ge source ED-TFET has maximum number of overlapping energy states for both OFF and ON states, hence, higher current in both states as compared to Si ED-TFET.

\begin{figure}
\center
\includegraphics[width=90mm,keepaspectratio]{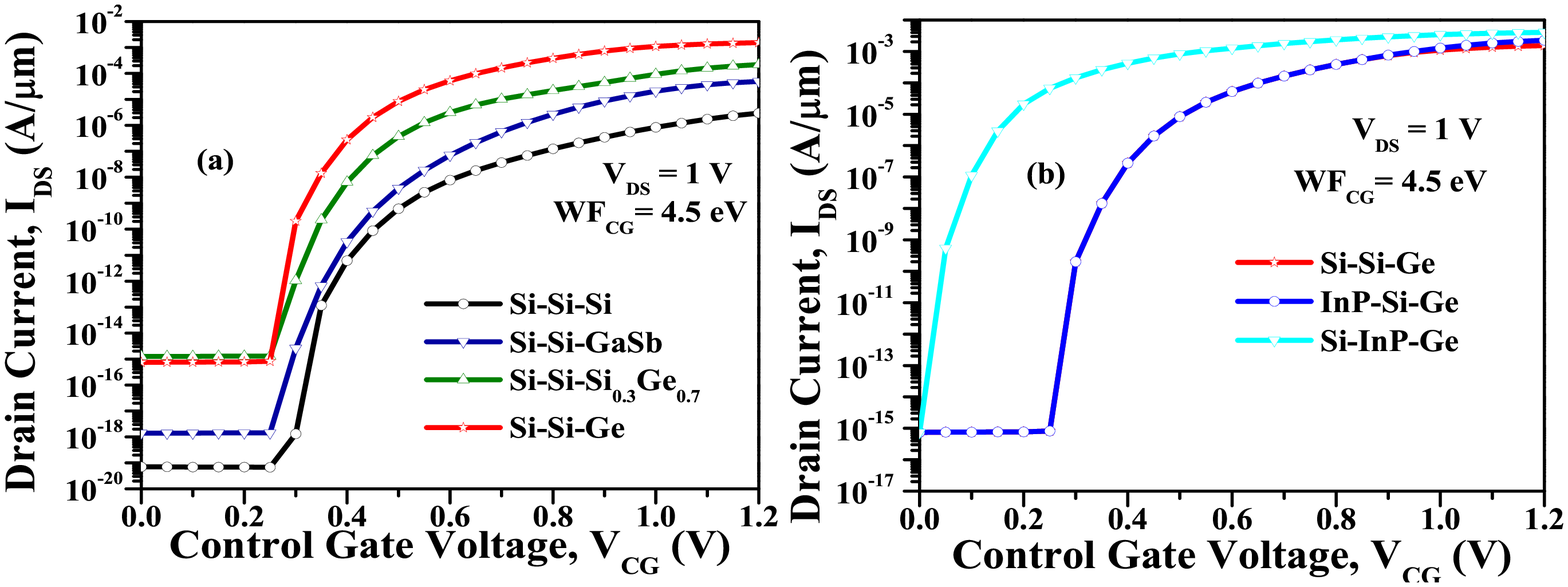}
\caption{Transfer characteristics of ED-TFET with various interfacing of group III-V and IV semiconductor materials (a) low band gap materials in source region, (b) high band gap materials in drain and channel region in place of silicon.}
\label{lbg}
\end{figure}

Fig.\ref{lbg} (a) and (b) show the transfer characteristics of ED-TFET with interfacing of semiconductor materials (group III-V and group IV) along with band gap engineering employed for heterojunctions at source-channel as well as at drain-channel junction with 4.5 eV control gate work function. From Fig.\ref{lbg} (a), one can observed that the ON current of ED-TFETs is increasing with the incorporation of low band gap materials like Germanium (Ge), Silicon-Germanium ($Si_{0.3}$$Ge_{0.7}$), and Gallium Antimonide (GaSb) as a source region material. The smaller source band-gap enables enhanced band-to-band tunneling rate, and therefore, high ON current but at the expense of large OFF current in comparison with silicon source. Amongst all low band gap materials in the source region, germanium offers high ON current with significantly low OFF current.

In Fig.\ref{lbg} (b), we have further optimized the ON and OFF state currents of ED-TFET with germanium source region by employing combinations of group III-V and IV materials (high band gap) to realize hetero drain-channel junction. Influence of large band gap materials than silicon (1.12 eV) like Indium Phosphide (1.35 eV) as drain and channel material is clearly visible on the device performance. Incorporation of high band gap materials in channel as well as in drain region helps in reduction of carrier recombination rate and in turn low $I_{OFF}$). In spite of that, it also results in an increase in energy barrier width at the drain and channel interface due to the increase in energy difference between conduction band edge of channel and valence band edge of drain which further decreases the reverse current, as a result, reduced ambipolar behavior.

\begin{figure} 
\center
\includegraphics[width=90mm,keepaspectratio]{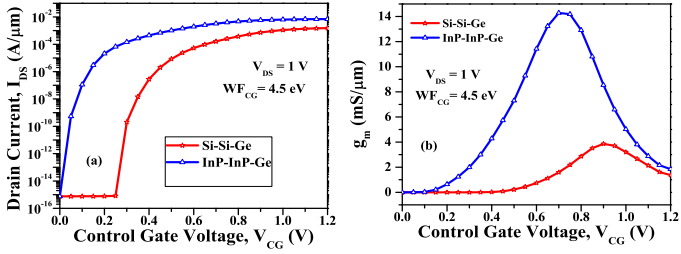}
\caption{Transfer characteristics comparison of ED-TFET with (a) high band gap (InP and GaAs) at drain as well as channel region, and (b) transconductance with control gate voltage variation with fixed $V_{DS}$ = 1V.}
\label{DCC}
\end{figure}

Fig.\ref{DCC} (a) show the comparison of transfer characteristics of ED-TFET  with low band gap germanium source along with that of high band gap InP drain-channel region.  It is confirmed that the ED-TFET exhibits further enhanced ON current with the inclusion of InP as drain as well as a channel region material. This is due to the high electron velocity of InP material. Hence, it can be concluded that the use of high band gap materials at drain as well as channel region in addition with source having low band gap materials may increase the ON-state current and lower the OFF state current  with the reduction in ambipolar behaviour of device. Fig.\ref{DCC} (b) show the transconductance variations corresponding to the two combination of hetero ED-TFET in Fig.\ref{DCC} (a) with control gate voltage variation. InP-InP-Ge ED-TFET  exhibits higher value transconductance due to the high value of ON-state current in respect of Si-Si-Ge ED-TFET.

\begin{table} 
\centering
\caption{Performance comparison of homo and heterojunction ED-TFETs}
\begin{center}
\begin{small}
\renewcommand{\arraystretch}{1}
\begin{tabular}{|l|l|l|l|}\hline
Parameters              & Si-Si-Si         & Si-Si-Ge        & InP-InP-Ge        \\ \hline
$I_{ON}$   mA/$\mu$m    &  0.0029          & 1.58            & 7                 \\ \hline
$I_{OFF}$  A/$\mu$m     &  7.00x$10^{-20}$ & 8.00x$10^{-16}$ & 7.47x$10^{-16}$   \\ \hline
Point $SS$ mV/dec       &  10.1            & 9.5             & 8.5               \\ \hline
\end{tabular}
\end{small}
\label{Table1}
\end{center}
\end{table}

Table-I shows the comparison of performance characteristics of heterojunction ED-TFETs, comprised of low band gap material (Ge) in source region and high band gap material (InP) in drain and channel regions. It can be seen that hetero ED-TFETs has superior performance in comparison with homo silicon ED-TFET. Therefore it confirmed the potential benefit of incorporation of low and high band gap materials along with interfacing of group III-V and IV materials for enhancing the performance of tunnel-FETs. Although InP-InP-Ge ED-TFET has paramount boost in performance characteristics yet the possible fabrication is need to be addressed. Hence Si-Si-Ge ED-TFET with significant performance is the preferred hetero ED-TFET in terms of reduced fabrication complexity. The experimental studies of TFETs based on Ge source have also revealed the potentiality of Ge material to be a best low band gap source material having superior performance at low operating voltages~\cite{Hwan}. Further, Ge having small lattice mismatch (4$\%$) with silicon in comparison with other low band gap materials, yields better interfacing with silicon~\cite{Goley}. The Si-Si-Ge ED-TFET outperforms when comparing the simulation results from the table with some recently developed hetero TFETs i.e., vertical InAs:Si nanowire heterojunction tunnel transistors ($I_{ON}$ $\sim$2.4 $\mu$A)~\cite{Bessire} and $p^{+}$SiGe:Si:$n^{+}$Si hetero TFET ($I_{ON}$ $\sim$0.42 mA) \cite{Kim}.

\begin{figure} 
\center
\includegraphics[width=90mm,keepaspectratio]{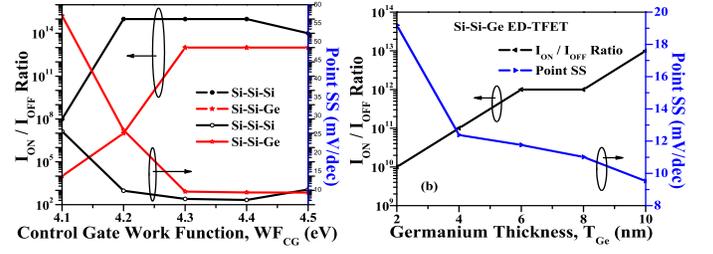}
\caption{(a) Comparison of transfer characteristics of silicon and germanium source ED-TFETs,  and (b) output characteristics of germanium source ED-TFET with gate voltage variations at fixed $V_{DS}$ = 1V.}
\label{Opt}
\end{figure}

Fig.\ref{Opt} (a) and (b) show the  performance optimization of Si-Si-Ge ED-TFET comparing with Si-Si-Si ED-TFET. Fig.\ref{Opt} (a) show the $I_{ON}$/$I_{OFF}$ ratio and point SS with variation in control gate work function for both the devices. The optimum value of control gate work function is 4.3 eV at which both devices have better performance. Fig.\ref{Opt} (b) show the $I_{ON}$/$I_{OFF}$ ratio and point SS of Si-Si-Ge ED-TFET with variation in thickness of germanium material in source region. It can be seen that both parameters are increasing as we decreases the thickness of germanium in source region. The thin epitaxial layer of germanium is easier to fabricate and interface with silicon through direct epitaxy method ~\cite{Goley}. Hence the optimum value of germanium thickness can be taken as 6 nm with less degradation in these parameters.  This further added an advantage in having simpler fabrication process with lower manufacturing cost.

\begin{figure} 
\center
\includegraphics[width=90mm,keepaspectratio]{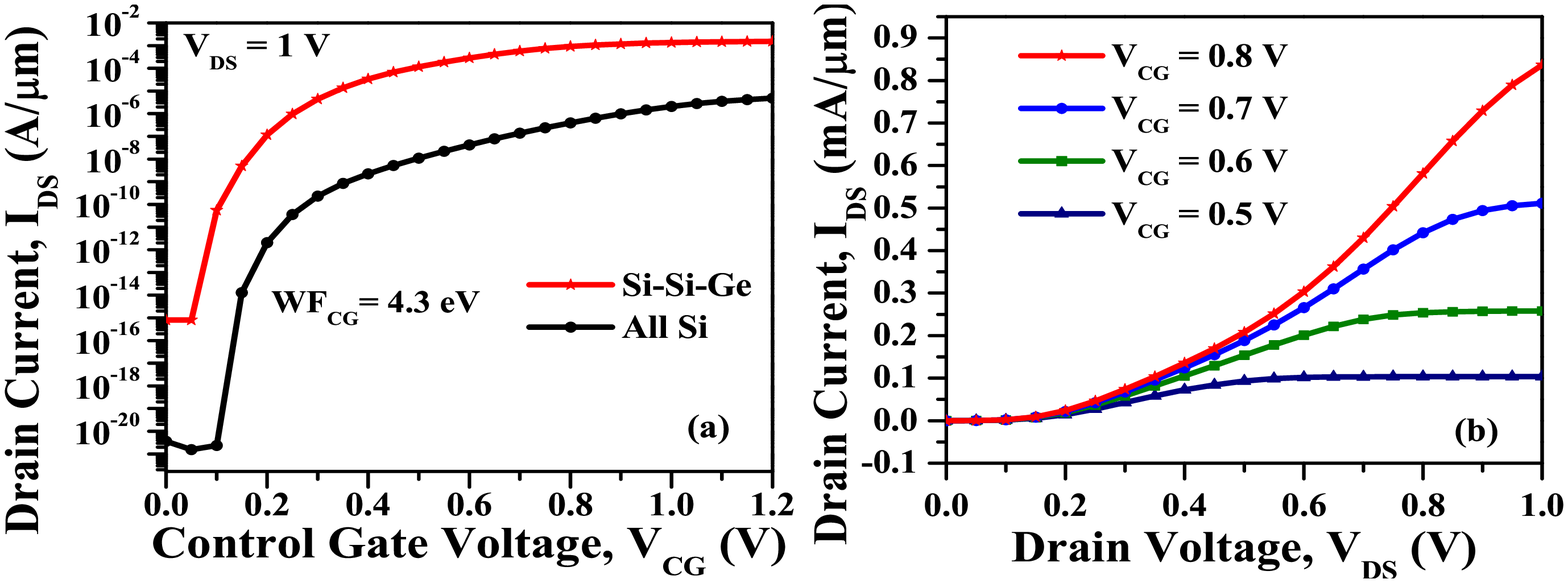}
\caption{(a) Transfer characteristics of silicon and germanium source ED-TFETs,  and (b) output characteristics of germanium source ED-TFET.}
\label{idvd_2}
\end{figure}

Fig.\ref{idvd_2} (a) shows the transfer characteristics of silicon and germanium source ED-TFETs. Amongst all the low band gap materials, germanium ensures higher ON current of 1.58 mA with point SS of 9.5 mV/dec and $I_{ON}$/$I_{OFF}$ ratio$\sim$$10^{13}$ in comparison with 0.0048 mA ON current, point SS of 7.5 mV/dec and $I_{ON}$/$I_{OFF}$ ratio$\sim$$10^{15}$ of Si ED-TFET at 4.3 eV control gate work function. Fig.\ref{idvd_2} (b) shows the output characteristics of ED-TFET with germanium source (Si-Si-Ge) for different values of $V_{GS}$. It can be inferred that in triode region as the drain voltage $V_{DS}$ increases the tunneling of carriers increases, and hence, $I_{ON}$ also increases. This establishes an exponential relationship with $I_{DS}$ due to drain induced barrier thinning~\cite{Sahu}. With further increase in $V_{DS}$, the tunneling width becomes less reliant on it, hence, causes the saturation of the drain current. The output characteristic of (Si-Si-Ge) ED-TFET exhibits more exponential $I_{DS}$-$V_{DS}$ relationship, and thereby more linearity in the operation.

\section{Analog performance}
\begin{figure} 
\center
\includegraphics[width=90mm,keepaspectratio]{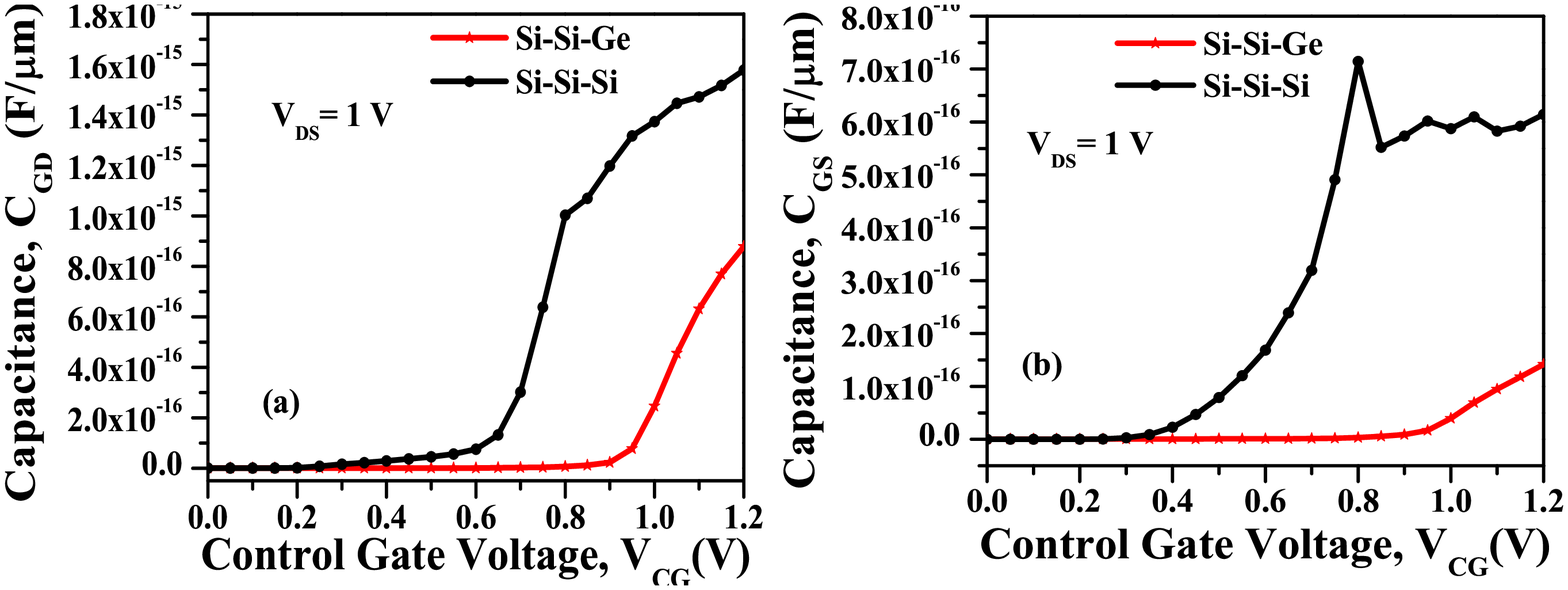}
\caption{(a) Gate to drain ($C_{GD}$), and (b) gate to source ($C_{GS}$) capacitances with variation in $V_{GS}$ for both Si-Si-Si and Si-Si-Ge ED-TFET.}
\label{Cap_Comp}
\end{figure}

\begin{figure} 
\center
\includegraphics[width=90mm,keepaspectratio]{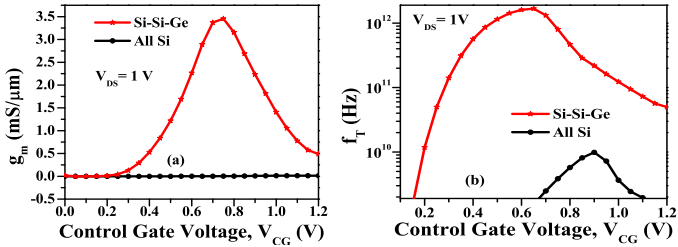}
\caption{(a) Transconductance ($g_{m}$) and (b) unity gain frequency ($f_{T}$) with variation in $V_{GS}$ for both Si-Si-Si and Si-Si-Ge ED-TFET.}
\label{analog}
\end{figure}

For the optimized germanium source ED-TFET, we have also studied the analog performance and compared with silicon source ED-TFET. Fig.\ref{Cap_Comp} shows the parasitic capacitances, (a) gate to drain capacitance ($C_{GD}$) and (b) gate to source capacitance ($C_{GS}$), extracted from the transfer characteristics for both Si and Ge source of ED-TFET as a function of $V_{GS}$ from the small signal ac analysis at a frequency of 1 GHz. It is observed that the values of these parasitic capacitances are very low in case of Ge source ED-TFET. The sum of these capacitances is referred as the total input/gate capacitance $C_{GG}$ ($C_{GS}$ + $C_{GD}$). Fig.\ref{analog} (a) and (b) show the transconductance ($g_{m}$) and unity gain frequency ($f_{T}$), respectively, as function of $V_{GS}$, extracted from the transfer characteristics of ED-TFET with both silicon and germanium source material at the same simulation conditions of capacitance extraction.

Transconductance $g_{m}$($\delta$$I_{D}$/$\delta$$V_{GS}$) describes the amplification capability of a device, higher value of $g_{m}$$\sim$3.37 mS/$\mu$m for Ge source ED-TFET ensures higher amplification, hence, it suits for analog applications. The unity gain frequency is expressed as $f_{T}$=$g_{m}$/2x$\pi$x$C_{GG}$, it is evident that as gate bias increases from subthreshold to saturation regime $f_{T}$ increases driven by higher transconductance, ON current and lower parasitic capacitances. Later, $f_{T}$ decreases after reaching a peak point (between the minimum gate-drain/source capacitance and peak of transconductance), due to the continuous increasing of the total gate-to-drain/source parasitic capacitances ($C_{GS}$ + $C_{GD}$) and the limiting value of $g_{m}$ because of mobility degradation with gate field. The hike in value of $f_{T}$$\sim$THz range of ED-TFET with Ge source enables its potentiality for RF applications.

\section{Conclusion}
A simplified fabrication process, low thermal budget, and less susceptibility to process variations makes ED-TFET a potential device for main stream VLSI applications. However, limited ON current restricts its suitability for such applications, therefore, heterojunctions based approach is investigated for different combinations of semiconductor materials employed for drain--channel--source regions. Is is observed that the high mobility and low band gap Ge as a source region material manifest a paramount boost in the derive current of ED-TFET. The analog and RF performance metrics i.e transconductance ($g_{m}$) and unity gain frequency $f_{T}$ have also improved for Ge source ED-TFET as compared to its counterpart Si source ED-TFET.


\begin{thebibliography}{99}
\bibitem{Riel} Ionescu, A.M., Riel, H., "Tunnel field-effect transistors as energy efficient  electronic switches," Nature, vol.479, no.7373, pp.329,337, Nov. 2011
\bibitem{Woo} Woo Young Choi et al, "Tunneling Field-Effect Transistors (TFETs) With Subthreshold Swing (SS) Less Than 60 mV/dec," Electron Device Letters, IEEE , vol.28, no.8, pp.743,745, Aug. 2007
\bibitem{Boucart} Boucart, K.; Ionescu, A.M., "Double-Gate Tunnel FET With High-k Gate Dielectric," Electron Devices, IEEE Transactions on , vol.54, no.7, pp.1725,1733, July 2007
\bibitem{Changhwan} Damrongplasit et al, "Study of Random Dopant Fluctuation Effects in Germanium-Source Tunnel FETs," Electron Devices, IEEE Transactions on , vol.58, no.10, pp.3541,3548, Oct. 2011
\bibitem{Sahu} Lahgere, et al;  "PVT-Aware Design of Dopingless Dynamically Configurable Tunnel FET," in Electron Devices, IEEE Transactions on , vol.62, no.8, pp.2404-2409, Aug. 2015
\bibitem{Marchi} De Marchi, et al; "Polarity control in double-gate, gate-all-around vertically stacked silicon nanowire FETs," in Electron Devices Meeting (IEDM), 2012 IEEE International , vol., no., pp.8.4.1-8.4.4, 10-13 Dec. 2012
\bibitem{Royer} Kazazis, D. et al, "Tunneling field-effect transistor with epitaxial junction in thin germanium-on-insulator," Applied Physics Letters, 94, 263508 (2009)
\bibitem{Toh} Toh et al,"Performance enhancement of n-channel impact-ionization metal-oxide-semiconductor transistor by strain engineering," Applied Physics Letters, 90, 023505 (2007)
\bibitem{Yoon} Ganapathi et al, "Analysis of InAs vertical and lateral band-to-band tunneling transistors: Leveraging vertical tunneling for improved performance", Applied Physics Letters, 97, 033504 (2010)
\bibitem{Ghosh} Asthana, P.K et al, "High-Speed and Low-Power Ultradeep-Submicrometer III-V Heterojunctionless Tunnel Field-Effect Transistor," in Electron Devices, IEEE Transactions on , vol.61, no.2, pp.479-486, Feb. 2014
\bibitem{CSahu} Sahu, C.; Singh, J., "Charge-Plasma Based Process Variation Immune Junctionless Transistor," Electron Device Letters, IEEE , vol.35, no.3, pp.411,413, March 2014
\bibitem{Chao} Wang C et al, "Self-aligned fabrication of 10 nm wide asymmetric trenches for Si/SiGe heterojunction tunneling field effect transistors using nanoimprint lithography, shadow evaporation, and etching", Journal of Vacuum Science and Technology B, 27, 2790-2794 (2009)
\bibitem{Zhou}  Zhou,G. et al, "InGaAs/InP Tunnel FETs With a Subthreshold Swing of 93 mV/dec and ION/IOFF Ratio Near 106 ," in Electron Device Letters, IEEE , vol.33, no.6, pp.782-784, June 2012
\bibitem{Atlas} Silvaco Int.,  Santa Clara, CA, USA. (2014). ATLAS Device Simulation Software. [Online]. Available: http://www.silvaco.com
\bibitem{Buca} Schmidt et al, "Line and Point Tunneling in Scaled Si/SiGe Heterostructure TFETs," in Electron Device Letters, IEEE , vol.35, no.7, pp.699-701, July 2014
\bibitem{Hwan} S H Kim et al, "Germanium-source tunnel field effect transistors with record high ION/IOFF," VLSI Technology, 2009 Symposium on , vol., no., pp.178,179, 16-18 June 2009
\bibitem{Goley} Goley, P.S.; Hudait, M.K., Germanium Based Field-Effect Transistors: Challenges and Opportunities. Materials 2014, 7, 2301-2339.
\bibitem{Bessire} Riel et al, "InAs-Si heterojunction nanowire tunnel diodes and tunnel FETs," in Electron Devices Meeting (IEDM), 2012 IEEE International , vol., no., pp.16.6.1-16.6.4, 10-13 Dec. 2012
\bibitem{Kim} Yoonmyung Lee et al, "Low-Power Circuit Analysis and Design Based on Heterojunction Tunneling Transistors (HETTs)," in Very Large Scale Integration (VLSI) Systems, IEEE Transactions on , vol.21, no.9, pp.1632-1643, Sept. 2013
\end{thebibliography}
\end{document}